\documentclass{article}

\usepackage{arxiv}

\usepackage[utf8]{inputenc}
\usepackage[T1]{fontenc}
\usepackage{hyperref}
\usepackage{url}
\usepackage{booktabs}
\usepackage{amsfonts}
\usepackage{amssymb}
\usepackage{amsmath}
\usepackage{nicefrac}
\usepackage{microtype}
\usepackage{graphicx}
\usepackage{cuted}
\graphicspath{{./}}

\title{OpenAaaS: An Open Agent-as-a-Service Framework for Distributed Materials-Informatics Research}

\author{
  Peng Kang \\
  National Key Laboratory of AI for Materials Science \\
  Beihang University \\
  Beijing, China \\
  \texttt{pengkang@buaa.edu.cn} \\
  \And
  Bixuan Li \\
  Tianmushan Laboratory\\
  Beihang University \\
  Beijing, China \\
  \texttt{buaa\_lbx@buaa.edu.cn} \\
  \And
  Xiaoya Huang \\
  Tianmushan Laboratory \\
  Beihang University \\
  Beijing, China \\
  \texttt{bhtb024@tmslab.cn} \\
  \And
  Shuo Shi \\
  Tianmushan Laboratory \\
  Beihang University \\
  Beijing, China \\
  \texttt{bht0062@tmslab.cn} \\
  \And
  Weiqiao Zhou \\
  Tianmushan Laboratory \\
  Beihang University \\
  Beijing, China \\
  \texttt{bht0080@tmslab.cn} \\
  \And
  Zhen Li \\
  National Key Laboratory of AI for Materials Science \\
  Beihang University \\
  Beijing, China \\
  \texttt{zli@buaa.edu.cn} \\
  \And
  Yu Liu\thanks{Corresponding author} \\
  National Key Laboratory of AI for Materials Science \\
  Beihang University \\
  Beijing, China \\
  \texttt{liuyucarbon@buaa.edu.cn} \\
  \And
  Lei Zheng\thanks{Corresponding author} \\
  National Key Laboratory of AI for Materials Science \\
  Beihang University \\
  Beijing, China \\
  \texttt{zhenglei@buaa.edu.cn}
}

\begin{document}
\maketitle

\begin{abstract}
The Materials Genome Initiative catalyzed the proliferation of centralized platforms---SaaS, PaaS, and IaaS---that aggregate computational and experimental resources for accelerated materials discovery. In parallel, breakthroughs in large language models (LLMs) and autonomous agents have created powerful new reasoning capabilities for scientific research. Yet a critical ``last mile'' problem remains: while we possess world-class models and vast repositories of materials data, we lack the organizational infrastructure to compose these capabilities securely across institutional boundaries. The development of structural and functional materials for harsh service environments---high-temperature alloys, radiation-resistant steels, corrosion-resistant coatings---remains characterized by long-term iteration, mechanistic complexity, and high domain expertise---demands that exceed both monolithic agent systems and traditional centralized platforms.

To address this gap, we propose \textbf{OpenAaaS}, an open-source hierarchical and distributed Agent-as-a-Service framework that enables organized multi-agent collaboration for intelligent materials design. OpenAaaS is built on a single foundational principle: \textit{code flows, data stays still}. A Master Agent plans and decomposes complex research tasks without requiring direct access to subordinate agents' managed data and computational resources. Sub-agents, deployed as near-data execution nodes, retain full sovereignty over local datasets, proprietary algorithms, and specialized hardware. This architecture guarantees that raw data never leaves its domain of origin while enabling cross-scale, cross-domain secure integration of previously isolated materials intelligence silos. We validate the framework through two representative case studies: (i) AlphaAgent, an evidence-grounded materials literature analysis executor that achieves 4.66/5.0 on deep analytical questions against single-pass RAG baselines; and (ii) an ultra-large-scale hexa-high-entropy alloy descriptor database service that demonstrates secure near-data execution and domain-specific scientific workflows under strict data-sovereignty constraints. OpenAaaS establishes a principled pathway toward ``organized research'' via agent collectives, offering a scalable foundation for next-generation materials intelligent design platforms. All source code is available at \url{https://github.com/Wolido/OpenAaaS}.
\end{abstract}


\section{Introduction}
\label{sec:intro}

\subsection{From the Materials Genome Initiative to Data-Driven Design}

The launch of the Materials Genome Initiative (MGI) in 2011 marked a paradigm shift in materials science, establishing the strategic vision that materials discovery and deployment could be accelerated by integrating computational tools, experimental data, and digital infrastructure into a unified innovation ecosystem~\cite{mgi2011}. In the decade that followed, this vision materialized through the construction of large-scale materials databases and integrated computational platforms. The Materials Project~\cite{jain2013,horton2025natmat}, AFLOW~\cite{curtarolo2012}, the Open Quantum Materials Database (OQMD)~\cite{kirklin2015}, and the NOMAD Laboratory~\cite{draxl2019} emerged as foundational pillars, aggregating millions of density functional theory (DFT) calculations, crystal structures, and thermodynamic properties into queryable web services. These platforms democratized access to high-fidelity computational data and catalyzed a transition from intuition-driven to data-driven materials design~\cite{tshitoyan2019,merchant2023}.

Concomitantly, the informatics infrastructure evolved through layered service models. Software-as-a-Service (SaaS) platforms provided web-based tools for property prediction and visualization. Platform-as-a-Service (PaaS) offerings enabled researchers to deploy custom machine learning (ML) workflows atop shared computational backends. Infrastructure-as-a-Service (IaaS) layers abstracted the physical hardware needed for first-principles simulations. Together, these architectures formed a vertically integrated stack that lowered the barrier to entry for computational materials research~\cite{saal2020,ramakrishna2019}.

\subsection{The Rise of Large Language Models and Scientific Agents}

The emergence of large language models (LLMs) introduced a disruptive new capability into this landscape. Beginning with GPT-3~\cite{brown2020} and accelerating through instruction-tuned variants such as InstructGPT~\cite{ouyang2022} and GPT-4~\cite{achiam2023}, LLMs demonstrated emergent abilities in reasoning, code generation, and domain-agnostic knowledge synthesis~\cite{wei2022}. In materials science specifically, LLMs have been applied to crystal structure generation~\cite{antunes2024}, molecular design~\cite{sanchez2018}, literature mining~\cite{tshitoyan2019,jiang2025npj}, and autonomous experimental planning~\cite{boiko2023,bran2024}. Systems such as Coscientist~\cite{boiko2023}, ChemCrow~\cite{bran2024}, and SciAgents~\cite{ghafarollahi2024} have shown that LLM-based agents can plan, execute, and validate multi-step scientific workflows with minimal human intervention.

This progress has catalyzed a broader architectural transition: from static software services to dynamic, reasoning-capable agents delivered as composable network services. The concept of Agent-as-a-Service (AaaS)---where autonomous agents are exposed through standardized APIs and discovered dynamically by client systems---has gained significant traction in both industry and academia~\cite{aaasan2025,wu2023autogen}. AaaS decouples the agent's cognitive capabilities from the underlying infrastructure, allowing specialized scientific agents to be developed, versioned, and consumed independently.

\subsection{The Unmet Challenge: The Last Mile of Cross-Organizational Materials R\&D}

Despite these advances, a critical gap remains for the development of structural and functional materials intended for harsh service environments---high-temperature alloys for turbine blades, radiation-resistant steels for nuclear reactors, corrosion-resistant coatings for marine infrastructure, and multifunctional materials for aerospace applications. The research and development (R\&D) of such materials exhibits four distinguishing characteristics that strain existing architectures:

\begin{enumerate}
\item \textbf{Long-term horizon:} Development cycles span years to decades, involving iterative loops of synthesis, characterization, modeling, and validation that exceed the context window and persistence of single-session agents.
\item \textbf{Mechanistic complexity:} Performance is governed by coupled multiscale phenomena---electronic structure, defect chemistry, microstructural evolution, and environmental degradation---that require cross-domain synthesis rather than isolated tool invocation.
\item \textbf{High domain expertise:} Each sub-problem demands specialized knowledge in thermodynamics, kinetics, mechanics, or electrochemistry. A general-purpose agent lacks the depth to reason reliably across all relevant disciplines.
\item \textbf{Sensitive data constraints:} Proprietary alloy compositions, unpublished experimental results, and restricted computational models are often subject to confidentiality agreements, export controls, or institutional firewalls that prohibit data migration to centralized platforms.
\end{enumerate}

Recent surveys of LLM applications in materials discovery~\cite{jiang2025npj,miret2025nmi} and the broader transition from ``AI for Science'' to ``Agentic Science''~\cite{wei2025agentic,gridach2025survey} have underscored that the bottleneck has shifted from model capability to \textit{capability accessibility}: the ability to discover, compose, and orchestrate domain-specific scientific services across organizational boundaries. What is needed is an architecture that supports \textit{organized multi-agent collaboration}---a collective of specialized, persistent, and self-improving agents that can plan, delegate, reflect, and iterate across institutional boundaries while respecting data sovereignty.

\subsection{OpenAaaS: Organized Research via Hierarchical Agent Collectives}

To meet this need, we introduce \textbf{OpenAaaS} (Open Agent-as-a-Service, \url{https://github.com/Wolido/OpenAaaS}), a hierarchical, distributed multi-agent framework designed specifically for secure, cross-domain integration of materials intelligence resources. OpenAaaS is guided by a single foundational principle: \textit{code flows, data stays still}. Rather than migrating raw data to a central location for analysis, OpenAaaS distributes agent execution nodes to the locations where data already reside, transmitting only task descriptions, intermediate reasoning artifacts, and final results across the network.

The architecture comprises three hierarchical layers:
\begin{itemize}
\item \textbf{Master Agent Layer:} General-purpose LLM agents (e.g.,\ Kimi CLI, Claude Code, Codex, or custom systems) that understand user intents, decompose complex research problems into sub-tasks, and orchestrate the invocation of remote capabilities.
\item \textbf{Network Hub (OpenAaaS Server):} A lightweight Rust-based indexing and routing layer that maintains a registry of available services, routes tasks to qualified nodes, manages authentication, and relays files---without ever accessing the raw scientific data processed by the nodes.
\item \textbf{Network Node (Agent Core):} Domain-specific execution nodes deployed locally at data sites---laboratories, instrument workstations, or computational clusters. Each node runs in an isolated Docker sandbox and exposes its local datasets, analysis scripts, and specialized hardware as composable services to the network.
\end{itemize}

This layered design yields three critical advantages for materials R\&D. First, \textbf{data sovereignty is preserved}: because sub-agents execute near the data, raw datasets never traverse organizational firewalls. Second, \textbf{cross-domain composability is achieved}: a Master Agent can discover and orchestrate capabilities from multiple independent nodes---a literature-analysis service at one institution, a high-entropy alloy database at another, and a first-principles compute cluster at a third---without requiring any of them to expose their underlying data. Third, \textbf{the framework is progressively extensible}: new nodes can join the network by registering their capabilities through a self-describing API contract, and existing agents can consume these capabilities without code modifications, thanks to compatibility with the Model Context Protocol (MCP)~\cite{anthropic2024mcp}.

Recent advances in physics-aware multimodal multi-agent AI for alloy design~\cite{ghafarollahi2025pnas} and agentic frameworks for computational chemistry~\cite{pham2025chemgraph} and atomistic simulation~\cite{vriza2026} have demonstrated the scientific potential of multi-agent collectives. However, these systems largely assume that agents operate within a shared trust boundary---either on the same server or within the same organizational network. They do not address the cross-organizational, data-sovereignty constraints that are pervasive in industrial and national-laboratory materials research. OpenAaaS fills this gap by providing the first hierarchical AaaS architecture that natively encodes data locality and security as first-class design constraints rather than afterthoughts.

\subsection{Contributions and Paper Organization}

The contributions of this work are fourfold:
\begin{enumerate}
\item We formalize the architectural requirements for hierarchical multi-agent systems in materials science, identifying the tension between cross-domain integration and data sovereignty as a central design constraint that existing frameworks do not address.
\item We present the design and open-source implementation of OpenAaaS, a three-tier framework that realizes these requirements through near-data execution, progressive capability discovery, and standardized agent protocols. The framework is released under the MIT license at \url{https://github.com/Wolido/OpenAaaS}.
\item We demonstrate the framework's effectiveness through two case studies: (i) AlphaAgent, a materials-literature analysis executor that grounds answers in validated retrieval evidence and outperforms single-pass RAG baselines; and (ii) an ultra-large-scale hexa-high-entropy alloy descriptor database service that demonstrates secure near-data execution and domain-specific scientific workflows under strict data-sovereignty constraints.
\item We discuss the broader implications of organized agent collectives for ``organized research'' in materials science, outlining a roadmap for the transition from isolated digital infrastructures to an interconnected Agentic Science network.
\end{enumerate}

The remainder of this paper is organized as follows. Section~\ref{sec:related} reviews related work in materials informatics platforms, scientific LLM agents, multi-agent systems, and secure data-sharing architectures. Section~\ref{sec:architecture} presents the OpenAaaS framework in detail, including its design philosophy, communication protocols, and security model. Sections~\ref{sec:case1} and~\ref{sec:case2} describe the two case studies. Section~\ref{sec:discussion} discusses limitations, comparisons with existing approaches, and future directions. Section~\ref{sec:conclusion} concludes.

\section{Related Work}
\label{sec:related}

\subsection{Materials Informatics Platforms}

The MGI ecosystem has produced a rich landscape of materials data platforms, each addressing specific aspects of the discovery pipeline. The Materials Project~\cite{jain2013,horton2025natmat} provides DFT-computed properties for over 150,000 inorganic compounds through a RESTful API, enabling high-throughput screening for batteries, catalysts, and photovoltaic materials. AFLOW~\cite{curtarolo2012} and its associated AFLOWlib repository emphasize automated DFT workflow management and thermodynamic property calculation. OQMD~\cite{kirklin2015} focuses on formation-energy accuracy and has been extensively used for training machine learning interatomic potentials. NOMAD~\cite{draxl2019} pursues a broader FAIR-data mandate, storing not only computational results but also the full provenance of simulation workflows to ensure reproducibility.

Despite their individual strengths, these platforms share a common architectural pattern: they are \textit{centralized data warehouses}. Users upload queries and download results; the raw data reside on the platform's servers. This pattern works well for open, non-proprietary datasets but becomes problematic when data are subject to confidentiality, regulatory constraints, or institutional policies that forbid external hosting~\cite{wilkinson2016,sears2024}. Recent work has explored blockchain-based and federated alternatives for secure materials data sharing~\cite{blockchain2024}, yet these approaches often introduce significant computational overhead and governance complexity that limit their practical adoption in everyday laboratory workflows.

\subsection{LLM Agents for Scientific Discovery}

The application of LLM agents to scientific research has progressed rapidly along a spectrum from narrow tools to broad autonomous systems~\cite{ren2025,tang2024,wei2025agentic,gridach2025survey}. At the tool end, systems such as LitLLM~\cite{agarwal2024} and Paper Copilot~\cite{lin2024} automate literature review through embedding-based search and LLM re-ranking. ChemCrow~\cite{bran2024} wraps an LLM with 18 expert-designed chemistry tools and has autonomously planned and executed chemical syntheses, including novel chromophores. Coscientist~\cite{boiko2023}, published in \textit{Nature}, demonstrated that GPT-4 could design and execute palladium-catalyzed cross-coupling reactions with minimal human supervision.

In materials science specifically, several recent agents have targeted computational workflows. MatClaw~\cite{liu2024matclaw} introduced a ``code-first'' paradigm in which the agent writes and executes Python code composing domain libraries (pymatgen, atomate2, DeePMD-kit) rather than calling predefined tools. This design overcomes the pipeline-boundedness of earlier agents and enables orchestration of heterogeneous multi-code workflows. HoneyComb~\cite{zhang2024honeycomb} demonstrated flexible LLM-based agents for materials science by leveraging high-quality domain knowledge bases and tool-calling capabilities. DREAMS~\cite{wang2025dreams} proposed a density functional theory based research engine for agentic materials simulation, integrating automated structure generation with DFT validation.

The recent review by Jiang et al.~\cite{jiang2025npj} in \textit{npj Computational Materials} and the perspective by Miret and Krishnan~\cite{miret2025nmi} in \textit{Nature Machine Intelligence} have both emphasized that the transformative potential of LLMs in materials science depends critically on their ability to interface with domain-specific tools, databases, and experimental infrastructure. However, these surveys also note that current systems are predominantly single-agent or operate within closed computational environments, limiting their applicability to industrial R\&D scenarios where data cannot be centralized.

At the far end of the autonomy spectrum, The AI Scientist~\cite{lu2024} generates research ideas, writes and runs code, produces figures, drafts papers, and simulates peer review for under \$15 per paper. Its successor, AI Scientist-v2~\cite{yamada2025}, produced the first entirely AI-generated paper accepted at an ICLR workshop. Google's AI Co-Scientist~\cite{gottweis2025} applies multi-agent collaboration to hypothesis generation and experimental design.

\subsection{Multi-Agent Systems and Orchestration}

The transition from single agents to multi-agent collectives has been driven by the recognition that complex scientific problems require distributed expertise. AutoGen~\cite{wu2023autogen} enables next-generation LLM applications via multi-agent conversation, where agents with different roles collaborate through structured dialogue. MetaGPT~\cite{hong2024} frames software development as a multi-agent collaborative process with specialized roles. CAMEL~\cite{li2023camel} explores ``mind'' exploration of LLM society through communicative agents.

In scientific domains, hierarchical multi-agent architectures have demonstrated particular promise. Ghafarollahi and Buehler~\cite{ghafarollahi2024,ghafarollahi2025pnas} showed that physics-aware multimodal multi-agent systems can navigate knowledge graphs and automate alloy design and discovery, achieving performance comparable to human experts. Vriza et al.~\cite{vriza2026} proposed multi-agent frameworks for end-to-end atomistic simulations, published in \textit{Digital Discovery}. The AaaS-AN framework~\cite{aaasan2025} introduced a service-oriented agent paradigm built upon the Role-Goal-Process-Service (RGPS) standard, enabling plug-and-play multi-agent systems with over 100 integrated agent services. InternAgent-1.5~\cite{feng2026internagent} presented a unified agentic framework for long-horizon autonomous scientific discovery.

However, existing multi-agent frameworks largely assume that agents operate within a shared trust boundary---either on the same server or within the same organizational network. They do not address the cross-organizational, data-sovereignty constraints that are pervasive in industrial and national-laboratory materials research. As noted in recent surveys of agentic AI for scientific discovery~\cite{wei2025agentic,gridach2025survey}, the development of secure, distributed multi-agent infrastructures remains an open challenge.

\subsection{Agent-as-a-Service and Protocol Standardization}

The concept of Agent-as-a-Service (AaaS) has emerged as a natural evolution of cloud computing paradigms~\cite{aaasan2025}. Rather than delivering static software functions via APIs, AaaS delivers autonomous reasoning capabilities that can interpret goals, plan actions, and adapt to context. The market for AI-as-a-Service is projected to reach \$43.3 billion by 2028, with intelligent agents representing the fastest-growing segment.

A critical enabler for AaaS interoperability is the Model Context Protocol (MCP), introduced by Anthropic in late 2024~\cite{anthropic2024mcp}. MCP standardizes how AI applications discover, describe, and invoke external tools through a client-server architecture with JSON-RPC messaging. Since its introduction, MCP has been adopted by OpenAI, Microsoft, Google, and Cloudflare, with over 20 million weekly SDK downloads~\cite{hasan2025}. MCP addresses the long-standing fragmentation in tool-calling interfaces across LLM frameworks (LangChain, AutoGen, CrewAI, LlamaIndex), establishing a ``USB-C for AI'' that allows any compliant client to connect to any compliant server without custom adapters~\cite{hou2025}.

OpenAaaS builds upon these advances but extends them in a crucial direction: from tool-level interoperability to \textit{organizational-level capability sharing}. While MCP standardizes how a single agent calls a single tool, OpenAaaS standardizes how an entire network of agents, data nodes, and computational services can be discovered, composed, and orchestrated across institutional boundaries.

\subsection{Security and Privacy in Materials Data}

The tension between data sharing and data protection is a persistent challenge in materials informatics~\cite{wilkinson2016,sears2024}. The FAIR principles (Findable, Accessible, Interoperable, Reusable)~\cite{wilkinson2016} provide a framework for open data stewardship, yet many high-value materials datasets---proprietary alloy compositions, failed experiments, radiation-damage microstructures---cannot be made openly accessible without compromising competitive advantage or national security. Blockchain-based approaches have been proposed for auditable, decentralized data sharing in materials genome engineering~\cite{blockchain2024}, but these systems face scalability limitations and high consensus overheads that render them impractical for real-time scientific workflows.

Near-data computing---processing data at its point of origin rather than migrating it to a central facility---offers an alternative path. The approach has been explored in high-performance computing and is increasingly relevant for AI-driven science, where TB-scale datasets and regulated sensitive samples make migration prohibitively expensive or legally impermissible. OpenAaaS operationalizes near-data computing at the agent level: each execution node processes its local data in place, transmitting only lightweight control messages and results across the network.

\section{The OpenAaaS Framework}
\label{sec:architecture}

\subsection{Design Philosophy}

OpenAaaS is designed around four core propositions that address the limitations of both centralized platforms and isolated agent systems:

\textbf{Data stays in situ; capabilities flow across nodes.} The real solution to data silos is not moving all data into one place---it is bringing analytical capabilities to where the data live. Every laboratory's accumulated datasets, algorithmic workflows, and domain expertise become composable capability units that any agent can directly invoke. Agents need not master the full depth of a field in advance; they simply discover, orchestrate, and invoke services from nodes around the world, continuously expanding their effective knowledge boundaries.

\textbf{Zero data migration eliminates migration loss.} Traditional solutions demand that data be aggregated into a centralized platform, inevitably introducing format conversion distortion, metadata loss, version divergence, and broken compliance audit chains. OpenAaaS builds no unified data warehouse. Data remains at its point of origin, preserved in its original storage format, directory structure, and access permissions. Analysis tasks arrive remotely as code and instructions; results are sent back. Raw data never leaves.

\textbf{Schema-free onboarding: raw formats are service capabilities.} We impose no upfront format requirements on data. JSON, CSV, Excel, MATLAB \texttt{.mat}, HDF5, vendor-specific binary formats from instruments---the local parsing and processing scripts on each node are themselves part of the network's capability. Agents invoke a combined ``parse + analyze'' service, rather than being required to pre-clean, standardize, or structure the data. Whatever format a lab already has, it is service-ready from day one.

\textbf{Near-data computing makes data-movement cost negligible.} Computation happens next to the data, not the other way around. The network only transmits task descriptions and execution results (KB--MB scale); raw data is processed on-site. For TB-scale datasets and regulated sensitive samples, this means no upload wait, no bandwidth bottleneck, and no outbound compliance review. The marginal cost of moving data approaches zero.

\subsection{Three-Tier Architecture}

\begin{figure*}[!t]
\centering
\includegraphics[width=0.95\textwidth]{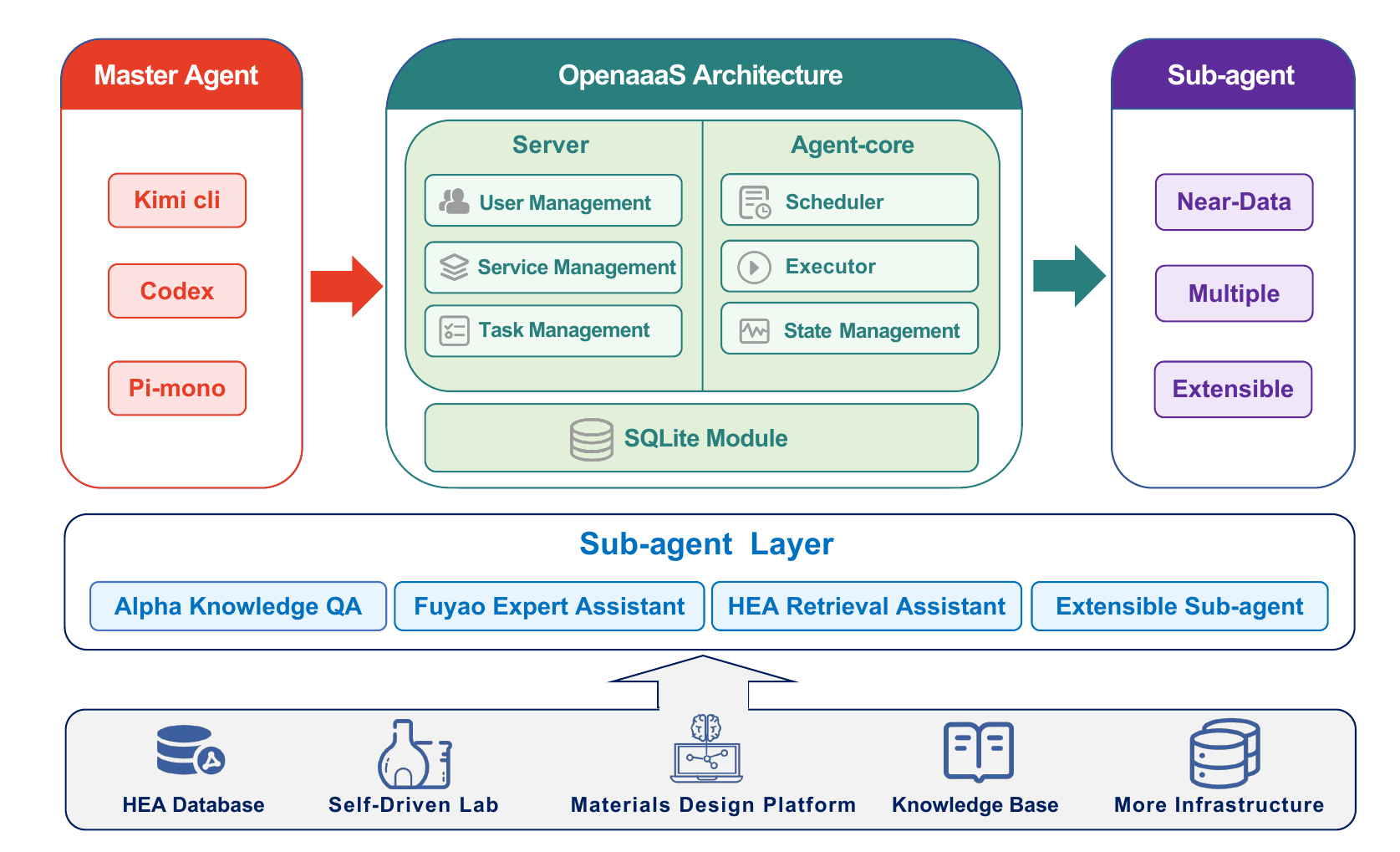}
\caption{The OpenAaaS hierarchical architecture. Master Agents (Kimi CLI, Codex, Pi-mono, or custom systems) interact with the OpenAaaS Network Hub (Server + Agent Core + SQLite Module), which routes tasks to specialized Sub-agents deployed as near-data execution nodes. Sub-agents expose local resources---HEA databases, self-driving labs, materials design platforms, knowledge bases---as composable network services without migrating raw data.}
\label{fig:architecture}
\end{figure*}

Figure~\ref{fig:architecture} illustrates the OpenAaaS architecture. The system is organized into three tiers that separate concerns of user interaction, network coordination, and domain execution. The complete implementation is available at \url{https://github.com/Wolido/OpenAaaS}.

\subsubsection{Tier 1: Master Agent Layer}

The Master Agent is the user's primary interface to the OpenAaaS network. It can be any general-purpose LLM agent that supports tool use or MCP connectivity: Kimi CLI, Claude Code, Codex, Pi-mono, or a custom-built system. The Master Agent's responsibilities are:
\begin{itemize}
\item \textbf{Intent understanding:} Parsing natural-language research goals into structured task specifications.
\item \textbf{Task decomposition:} Breaking complex, multi-step materials problems into sub-tasks that can be delegated to specialized network nodes.
\item \textbf{Capability discovery:} Querying the OpenAaaS Server for available services, filtering by domain, capacity, and trust constraints.
\item \textbf{Orchestration:} Submitting sub-tasks to appropriate nodes, monitoring execution status, and synthesizing results into coherent answers or reports.
\item \textbf{Reflection and replanning:} When sub-task results are insufficient or contradictory, the Master Agent can reformulate queries, select alternative nodes, or request deeper analysis.
\end{itemize}

Critically, the Master Agent \textit{never requires direct access to the data managed by subordinate nodes}. It reasons over task descriptions, service metadata, and returned results---not over raw experimental measurements, proprietary compositions, or restricted simulation outputs. This separation is what enables cross-organizational orchestration without violating data sovereignty.

\subsubsection{Tier 2: Network Hub (OpenAaaS Server)}

The Network Hub is a lightweight, single-binary HTTP server implemented in Rust, with an embedded SQLite database. It serves four functions:

\textbf{Service registration.} When a new Agent Core node joins the network, it registers its capabilities with the Server, providing a service description that includes supported request types, required input formats, output schemas, evidence levels, and example usage. This description is self-contained: an agent reading it can understand how to invoke the service without external documentation.

\textbf{Task routing.} The Server maintains a task queue and routes pending tasks to qualified nodes based on service type, node capacity, and heartbeat status. It supports load balancing across multiple nodes offering the same capability.

\textbf{Node heartbeat.} Nodes send periodic heartbeat messages reporting their current load, capacity, and health. The Server marks nodes as offline when heartbeat messages time out and automatically migrates queued tasks to alternative nodes.

\textbf{File relay.} The Server relays task input files from clients to nodes and result files from nodes back to clients. File sizes are bounded (default 50~MB), and storage is temporary (default 7-day retention). Raw scientific datasets are \textit{never} stored on the Server; only task descriptions, small input artifacts, and result files transit through it.

The Server requires only outbound HTTP connectivity from nodes, making it compatible with laboratory firewall configurations that block inbound connections. Nodes self-organize through reverse polling: they periodically query the Server for tasks rather than exposing listening ports themselves.

\subsubsection{Tier 3: Network Node (Agent Core)}

The Agent Core is the execution engine deployed at data sites. Implemented in Rust as a single binary, it:
\begin{itemize}
\item Registers with the Server using a one-time registration token.
\item Polls the Server for tasks assigned to its registered service.
\item Creates an isolated Docker container for each task, mounting a local workspace with task instructions and input files.
\item Executes the task inside the container sandbox, with full access to local datasets, analysis scripts, and specialized hardware (GPUs, instrument controllers).
\item Reports results and output files back to the Server upon completion.
\end{itemize}

Each task runs in an independent container with configurable resource limits (CPU, memory, timeout). This isolation ensures that a failed or malicious task cannot compromise the host system or other concurrent tasks. The container image is user-defined: a node can use a standard Python environment, a domain-specific image with VASP and pymatgen, or a custom executor with proprietary analysis code.

The Agent Core's near-data execution model means that raw datasets are accessed through local filesystem mounts or internal network connections inside the container---never through the public Server. For a node managing a 10~TB high-entropy alloy database, the database files remain on the node's local storage; only the query results (typically KB--MB) are returned to the client.

\subsection{Progressive Capability Discovery}

A key challenge in multi-agent networks is avoiding context overflow: if every service registration includes its full documentation, the Master Agent's context window is rapidly exhausted. OpenAaaS addresses this through progressive disclosure, inspired by the SKILL.md design pattern:

\begin{itemize}
\item \textbf{Stage 1 (Lightweight summary):} The Server returns a compact list of available services, including only name, domain tag, one-line description, and current capacity. This allows the Master Agent to filter candidates without loading detailed documentation.
\item \textbf{Stage 2 (On-demand usage):} When the Master Agent identifies a candidate service, it requests the full usage documentation, including input schemas, example prompts, output formats, and error handling. This detailed information is loaded only for the services under active consideration.
\item \textbf{Stage 3 (Interactive refinement):} For complex tasks, the Master Agent can submit a preliminary prompt to the service and receive clarifying questions or parameter suggestions, enabling iterative task refinement before full execution.
\end{itemize}

This three-stage design protects the Master Agent's reasoning capacity while ensuring that it has access to the precise information needed to construct valid task requests.

\subsection{MCP Compatibility and Client Extensions}

OpenAaaS is designed for broad client compatibility. Through the \texttt{openaaas-mcp-adapter}, any MCP-compatible client---Claude Desktop, Cursor, Cline, or custom MCP hosts---can connect to the OpenAaaS network with a single configuration entry. The adapter exposes 14 standard tools that map directly to OpenAaaS Server APIs.

For non-MCP agents, OpenAaaS provides dedicated plugins for Kimi CLI (Python-based) and Pi-mono (TypeScript-based). These plugins implement the same standard workflow: set server URL, register, browse services, get usage, submit task, query result, download output. All client extensions follow the progressive disclosure principle, ensuring consistent behavior across agent types. Client extensions and usage examples are documented in the repository at \url{https://github.com/Wolido/OpenAaaS}.

\subsection{Security Model}

OpenAaaS employs a defense-in-depth security model that addresses threats at the network, server, node, and container levels:

\textbf{Network layer.} All communication between clients, Server, and nodes uses HTTPS. Nodes only make outbound HTTP connections; no inbound ports are required. This ``reverse connection'' design ensures that laboratory firewalls need not be opened for external access.

\textbf{Authentication.} The Server uses API-key-based authentication with HMAC-SHA256 signatures. Each client and node receives a unique API key upon registration. Admin operations (creating services, viewing all tasks) require a separate admin API key.

\textbf{Node isolation.} Each Agent Core node is independently registered and authenticated. A compromised node cannot impersonate another node because API keys are service-specific and cryptographically bound to registration tokens.

\textbf{Container sandboxing.} Task execution occurs inside Docker containers with no network access to the host's internal systems (unless explicitly configured via mount points). Container images, resource limits, and timeout policies are controlled by the node operator, not by the client or Server.

\textbf{Data provenance.} The Server logs all task submissions, assignments, completions, and file transfers with timestamps and actor IDs. This audit trail supports compliance requirements without exposing the scientific content of the tasks or results.

Together, these mechanisms ensure that OpenAaaS can operate across trust boundaries---connecting academic, industrial, and governmental materials research nodes---while respecting the confidentiality and integrity constraints of each domain.

\section{Case Study I: AlphaAgent for Evidence-Grounded Materials Literature Analysis}
\label{sec:case1}

\subsection{Domain Motivation and Executor Role}

This section illustrates OpenAaaS through a concrete application example: a materials-science literature-analysis executor adapted from AlphaAgent~\cite{alphaagent2026}. In this example, OpenAaaS provides the agent framework, whereas the AlphaAgent executor implements a domain-specific workflow for scientific question answering and deep literature-report generation. The purpose of this case study is therefore to show what a specialized scientific executor contributes within the OpenAaaS framework. Such a domain-specific workflow is needed because materials-science conclusions are highly context-dependent. A claim about alloy strengthening, phase stability, or processing response may depend on composition, thermal history, characterization method, and testing condition. A general retrieval-augmented generation (RAG) pipeline may locate relevant-looking passages, but lexical or topical relevance alone does not guarantee that the selected evidence matches the material system, experimental context, mechanistic focus, and property target required for a defensible answer~\cite{lewis2020retrieval,izacard2021leveraging}. The executor addresses this gap by treating literature analysis as a controlled evidence procedure rather than as a single retrieve-then-generate step.

\subsection{Scientific Execution Contract}

The AlphaAgent executor exposes a scientific execution contract that specifies supported request types, allowed evidence sources, required intermediate artifacts, evidence levels, and returnable outputs. This contract is separate from the generic OpenAaaS service interface: OpenAaaS standardizes task delivery and execution, whereas the AlphaAgent executor defines the materials-literature rules needed to judge evidence validity. Table~\ref{tab:scientific_contract} summarizes this separation between platform-level execution and domain-level evidential control. After receiving a request, the executor evaluates whether the requested answer or report can be supported by an evidence chain that matches the material system, property target, mechanistic focus, and output type. The contract therefore maps each supported request mode, including grounded question answering, targeted paper reading, and cross-paper synthesis, to the minimum evidence level required for that mode. Lightweight answers may be grounded in validated retrieval snippets, whereas structured reports require document-level paper analysis before report generation.

\begin{table}[htbp]
\centering
\label{tab:scientific_contract}
\par\footnotesize
TABLE~\thetable\\
\textsc{Scientific execution contract for the AlphaAgent materials-literature executor}
\par\vspace{0.6em}
\begin{tabular}{@{}p{0.20\textwidth}p{0.72\textwidth}@{}}
\toprule
Contract element & Executor-level specification \\
\midrule
Supported requests & Retrieval-grounded scientific question answering, targeted paper reading, and cross-paper synthesis, optionally constrained by material system, processing route, property target, mechanistic focus. \\
Evidence sources & Retrieval is limited to the curated materials-literature index. Document-level analysis is performed only on papers selected on the basis of validated retrieval evidence. \\
Evidence level & Lightweight answers may use validated retrieval snippets, whereas structured reports require document-level paper analysis before report generation.\\
Retrieval behavior & The executor preserves materials-specific entities during intent construction and performs bounded query reformulation when evidence is insufficient or misaligned. \\
Intermediate state & Rewritten intents, retrieval attempts, candidate evidence, validated evidence, paper metadata, document references, citations, and failure records are retained. \\
Analytical output & Source-grounded answers, single-paper analytical reports, and cross-paper syntheses with explicit evidence scope. \\
Validity checks & Outputs are returned only when evidence alignment, paper availability, report-schema validation, and failure-aware synthesis meet the requirements of the requested output type. \\
\bottomrule
\end{tabular}
\par\vspace{0.8em}
\end{table}

\begin{figure*}[!t]
\centering
\includegraphics[width=0.92\textwidth]{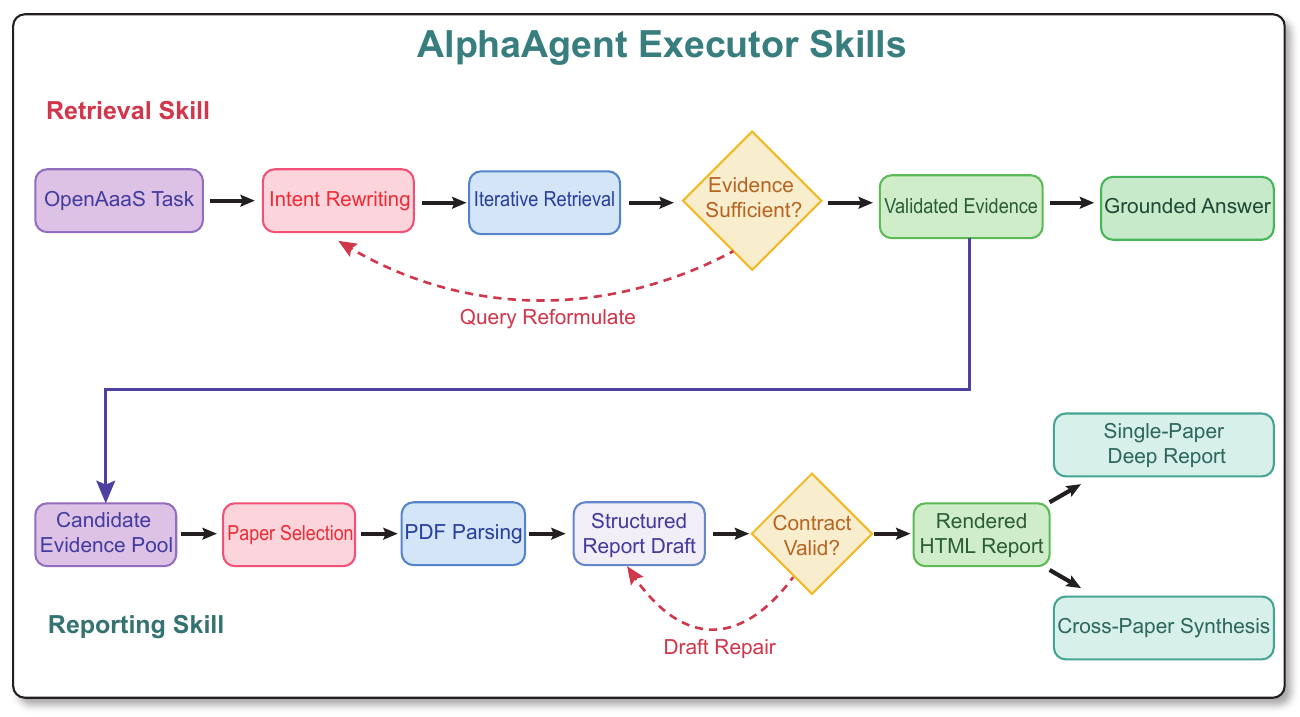}
\caption{Evidence-grounded skill composition of the AlphaAgent executor within OpenAaaS. The retrieval skill converts an OpenAaaS task into validated evidence through intent rewriting, iterative retrieval, evidence-sufficiency checking, and bounded query reformulation. Validated evidence supports grounded answering and, when deep reporting is requested, enters the reporting skill for paper selection, PDF parsing, structured report drafting, contract validation, and HTML rendering. The final report connects a question-centered cross-paper synthesis with linked single-paper deep reports, with feedback loops for query reformulation and draft repair.}
\label{fig:workflow_v2}
\end{figure*}

\subsection{Skill Composition and Evidence Handoff}

Figure~\ref{fig:workflow_v2} summarizes the executor as two coupled domain skills: retrieval-grounded question answering and deep literature reporting. The retrieval skill serves as the executor's RAG-style retrieval layer: it rewrites natural-language requests into English search intents, queries the curated materials-literature index, and returns candidate snippets and paper metadata for evidence validation. During intent rewriting, it preserves materials-specific entities, including alloy designations, phase names, heat-treatment conditions, property terms, and characterization methods. This preservation is important because small terminology changes can substantially shift the literature scope, especially in alloy systems where composition and processing route determine which mechanisms are relevant. After each retrieval attempt, the executor checks whether the returned snippets and metadata align with the requested material system, processing condition, property target, and mechanistic focus. When evidence is insufficient or misaligned, the executor performs bounded reformulation and updates the retrieval candidates. It then selects the validated evidence set that can support the requested output. In this way, evidence selection becomes an explicit control stage rather than an implicit consequence of the first retrieval result.

The reporting skill starts from the validated evidence state produced by the retrieval skill. It uses this state to select papers, parse PDFs, and generate structured report drafts. The rendered HTML report has two linked levels. The cross-paper synthesis is the question-centered layer: it expands the original request into multiple analytical angles and answers them using evidence from the selected literature. The single-paper deep reports are the paper-level layer, recording key results, mechanism analysis, innovations, evidence scope, and conclusions for the papers that support the synthesis. This structure gives researchers a navigable deep-reading artifact in which a synthesized answer and its detailed paper-level evidence remain connected. The same workflow also contains two control loops: weak retrieval evidence leads to query reformulation, and an invalid report draft leads to draft repair. These controls keep the answer, synthesis, and paper-level reports tied to the retrieval decisions, selected papers, and validation checks that produced them.

\subsection{Executor-Level Evaluation Against Baselines}

We use the AlphaAgent evaluation from the companion study~\cite{alphaagent2026} as an executor-level case study, comparing AlphaAgent with a controlled single-pass RAG baseline and two general-purpose model baselines, GPT-5.5 and Kimi-K2.6. The benchmark contains 40 metallurgical materials questions, including 20 deep analytical questions and 20 general questions. For the controlled RAG comparison, AlphaAgent and the single-pass baseline RAG answered these questions under the same retrieval setting: the same underlying model, the same retrieval scale, and the same literature index of more than 300,000 papers from the Journal Citation Reports \textit{Metallurgy \& Metallurgical Engineering} category. Table~\ref{tab:case_results_v2} summarizes the controlled RAG baseline and the general-model reference results. AlphaAgent achieves the highest average score on both task types. Its advantage is largest on deep analytical questions, where single-pass RAG is more vulnerable to retrieval drift, while general models tend to provide broad explanations that lack the mechanistic depth needed for materials-specific analysis.

\begin{table}[t]
\centering
\caption{Case-Study Results from AlphaAgent Evaluation}
\label{tab:case_results_v2}
\begin{tabular}{@{}lcc@{}}
\toprule
System & \begin{tabular}[c]{@{}c@{}}Deep Analytical\\Questions\end{tabular} & \begin{tabular}[c]{@{}c@{}}General\\Questions\end{tabular} \\
\midrule
AlphaAgent executor & \textbf{4.66} & \textbf{4.46} \\
GPT-5.5 & 4.05 & 3.96 \\
Kimi-K2.6 & 3.96 & 4.08 \\
Single-pass baseline RAG & 2.67 & 2.58 \\
\bottomrule
\end{tabular}
\end{table}

The deep analytical questions expose two different baseline failure modes. Single-pass RAG often suffers from retrieval drift: superficially similar papers or passages are retrieved, but the evidence does not match the requested mechanism, property target, or testing context. General models such as GPT-5.5 and Kimi-K2.6 are less tied to a specific retrieved evidence set, but their answers tend to remain broad when the question requires detailed mechanism analysis and paper-level support. AlphaAgent addresses these issues by preserving materials-specific entities during intent construction and validating whether retrieved evidence matches the analytical objective. It then uses the validated evidence to organize the answer around mechanisms, property targets, experimental context, and paper-level support, which turns matched retrieval into a deeper materials-specific analysis. Together, these controls explain why AlphaAgent is less vulnerable to retrieval drift and more effective on questions that require evidence-aligned mechanistic reasoning.

\subsection{Framework Boundary and Domain Reliability}

This case shows that OpenAaaS can separate the general task substrate from the evidence discipline required by a specific scientific domain. OpenAaaS provides a uniform way to host and invoke executors, while the materials-literature AlphaAgent executor defines the retrieval, validation, and reporting operations that make outputs trustworthy in its domain. Different scientific services can share the same framework while implementing different evidence contracts. For materials science, the executor must respect constraints on composition, processing history, phase structure, experimental method, and evidence provenance. Embedding these constraints at the executor level turns the service into a controlled scientific workflow, where intermediate evidence is preserved and failure modes are easier to localize. This is the broader implication of AlphaAgent for OpenAaaS: the shared framework provides a reusable task substrate, while domain reliability is achieved by encoding scientific evidence rules inside each executor.

\section{Case Study II: Ultra-Large-Scale Hexa-High-Entropy Alloy Descriptor Database}
\label{sec:case2}

\subsection{Domain Motivation and Executor Role}

This section illustrates the OpenAaaS architecture through a second concrete application: an executor for the ultra-large-scale hexa-high-entropy-alloy (HEA) descriptor database~\cite{xiaoya2026}, hereafter referred to as the HEA-Executor. Within the OpenAaaS ecosystem, the HEA-Executor operates as a domain-specific sub-agent deployed at a near-data execution node. OpenAaaS furnishes the distributed task-scheduling and service-oriented execution substrate; the HEA-Executor implements an automated scientific workflow that integrates high-dimensional materials-data querying, machine-learning modeling, and structured analytical-report generation over datasets spanning tens of billions of candidate records.

The rationale for constructing such a domain-specific executor stems from two fundamental characteristics of the HEA design space. First, the relationship between elemental composition, microstructural phase, and mechanical properties in six-component HEA systems is extraordinarily high-dimensional, strongly nonlinear, and governed by coupled thermodynamic and kinetic constraints. General-purpose large language models, which rely predominantly on statistical pattern matching over textual corpora, lack verifiable, composition-resolved associations with specific performance metrics. Conventional relational databases, by contrast, support only rudimentary conditional retrieval and are ill-equipped for high-dimensional descriptor correlation analysis, cross-system batch exploration, or physics-informed model inference. Consequently, intelligent exploration of this design space demands an executor that tightly integrates materials-science domain knowledge, machine-learning predictive models, and database-orchestration logic into a unified, controllable pipeline.

\begin{figure}[htbp]
\centering
\includegraphics[width=0.95\textwidth]{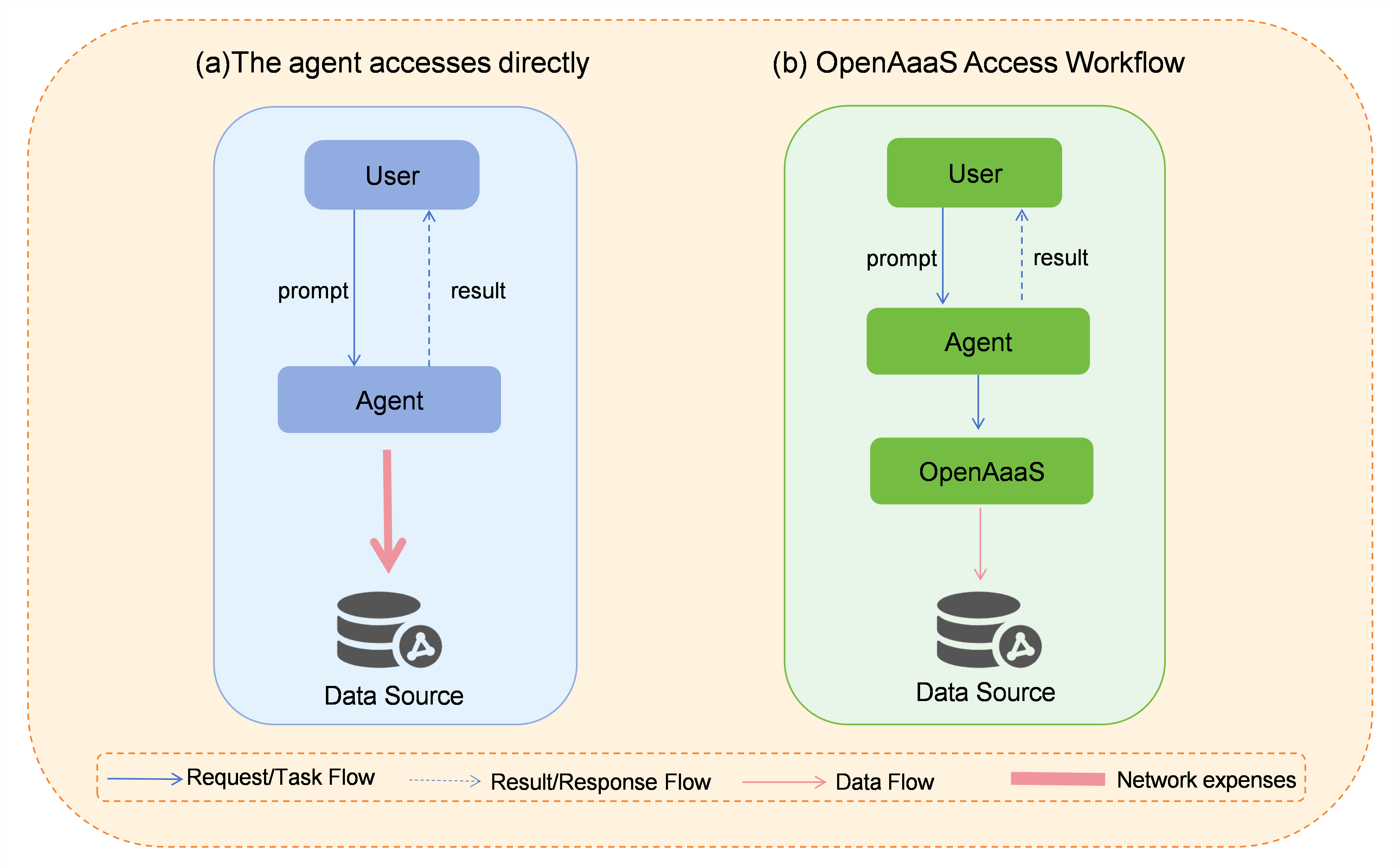}
\caption{Data-access paradigms for the HEA descriptor database. (a)~Direct-agent access follows a download-and-process-locally model that imposes prohibitive bandwidth and storage demands for terabyte-scale datasets. (b)~OpenAaaS-mediated retrieval deploys task execution at the data node, transmitting only lightweight task descriptions and filtered results across the network.}
\label{fig:case2_access}
\end{figure}

Second, the physical scale of the underlying dataset renders the conventional ``download-and-process-locally'' paradigm infeasible. The database comprises six-component HEAs drawn from a 15-element palette (Al, Co, Cr, Cu, Fe, Mn, Mo, Nb, Ni, Ti, V, W, Zr, Ta, Hf), spanning 5,005 unique elemental combinations and approximately $5.4 \times 10^{10}$ candidate compositions, each characterized by a 194-dimensional descriptor vector. The total data volume is approximately 17.4~TB. Under a direct-access model, constrained by standard institutional egress bandwidth of 100~Mbps, a full-dataset transfer would require more than 17.7 days---a latency that renders interactive design workflows impractical. As illustrated in Figure~\ref{fig:case2_access}, OpenAaaS replaces this data-migration bottleneck with a Computation-Near-Data strategy: the HEA-Executor performs filtering, model inference, and report generation at the data side, returning only essential results and structured summaries. This approach eliminates bulk data transfer, reduces client-side resource requirements, and preserves the data-sovereignty guarantees that are architecturally central to OpenAaaS.

\subsection{Executor Protocol Rules}

The database underlying the HEA-Executor consists primarily of computationally derived HEA descriptor data, supplemented by compressive-plasticity predictions from pre-trained machine-learning models; no proprietary experimental test data are included. To ensure the accessibility, schedulability, and security of ultra-large-scale materials databases within the OpenAaaS framework, the HEA-Executor defines five system-level protocol rules that govern every interaction with the node. These rules are enforced at the executor level, independent of the generic OpenAaaS service interface, and collectively encode the principle that \textit{code flows, data stays still} into operational constraints specific to terabyte-scale scientific databases.

\textbf{(1) Remote-Execution-First Protocol.} The HEA-Executor does not permit users or the Master Agent to download, copy, or transfer the raw database in full. All querying, filtering, statistical analysis, and model-inference tasks must be submitted as structured task instructions. The executor completes computations at the data side and returns only result data, statistical summaries, or structured reports. This protocol mandates Computation Near Data and prevents large-scale transfer of terabyte-scale databases over wide-area networks.

\textbf{(2) Minimal Data Return Protocol.} The executor returns only the minimal result set required to satisfy the task, rather than intermediate data arrays or the complete descriptor space. For example, when a user requests performance predictions for a specific composition range, the HEA-Executor outputs only the statistical information, filtered results, or model inference outcomes for the corresponding samples, without exposing the full database contents. This rule minimizes network bandwidth utilization and mitigates the risk of incremental data reconstruction by repeated querying.

\textbf{(3) Database Isolation Protocol.} Users, external agents, and third-party services are prohibited from directly accessing underlying database files, storage structures, or raw indexes. The HEA-Executor performs task parsing and scheduling through a unified interface, achieving logical isolation between the user side and the database side. All data access must be completed through protocolized interfaces; direct filesystem access to data sources is strictly forbidden.

\textbf{(4) Restricted Descriptor Exposure Protocol.} By default, the executor does not return the complete 194-dimensional descriptor set. Only the feature subsets, dimensionality-reduction results, or aggregate statistical features necessary for the specific task are emitted. Information that could expose the global database structure---including complete composition-space mappings, large-scale index distributions, and batch descriptor matrices---is restricted by default and released only under explicit administrative authorization.

\textbf{(5) Multi-User Shared Computing Protocol.} The HEA-Executor supports concurrent multi-user task scheduling, yet task contexts, cached results, and intermediate data are strictly isolated among different users. The system shares only underlying computational resources (CPU, memory, I/O bandwidth), not user task data or partial results, thereby balancing open-service capability with data security.

\begin{table}[htbp]
\centering
\caption{System-level protocol rules enforced by the HEA-Executor}
\label{tab:protocol_rules}
\begin{tabular}{@{}clp{7.2cm}@{}}
\toprule
No. & Protocol Name & Description \\
\midrule
1 & Remote-Execution First & Compute-at-data-side; remote execution; avoid full-dataset transfer; eliminate bandwidth dependence \\
2 & Minimal Data Return & Return only minimal result sets; on-demand emission; reduce redundant data exposure; control transfer volume \\
3 & Database Isolation & Interface-based access only; logical isolation between user and storage layers; direct access prohibited \\
4 & Restricted Descriptor Exposure & Feature-subset emission by default; prevent structural leakage; dimensionality-reduction results where applicable \\
5 & Multi-User Shared Computing & Concurrent scheduling with resource sharing; strict task-level isolation of intermediate data and cached results \\
\bottomrule
\end{tabular}
\end{table}

Together, these five protocol rules transform the HEA-Executor from a passive data endpoint into an active, policy-enforcing scientific gateway that provides secure access, remote inference, and structured computation over ultra-large-scale HEA databases under the OpenAaaS architecture.

\subsection{Executor Operational Logic}

The HEA-Executor reconstructs six-component HEA data analysis as a \textit{controlled multi-stage computational pipeline} rather than as a conventional single-query-retrieval operation. At the apex of this pipeline, a domain-specific coordinator (the \texttt{hea-master} agent) orchestrates three downstream sub-agents---responsible for data access, machine-learning modeling, and structured report generation, respectively. This layered design ensures end-to-end traceability from the raw database to the final deliverable: the output of each stage constitutes the validated input for the next, and intermediate computational states are explicitly preserved, enabling rapid localization and isolation of failure modes at module boundaries. By embedding materials-science-specific constraints---such as the permutation invariance of elemental combinations, the continuity of compositional space, and the nonlinear correlations among descriptors---into the operational specifications of each sub-agent, the executor elevates generic database querying to a controlled scientific workflow tailored for HEA design.

\begin{figure*}[!t]
\centering
\includegraphics[width=0.95\textwidth]{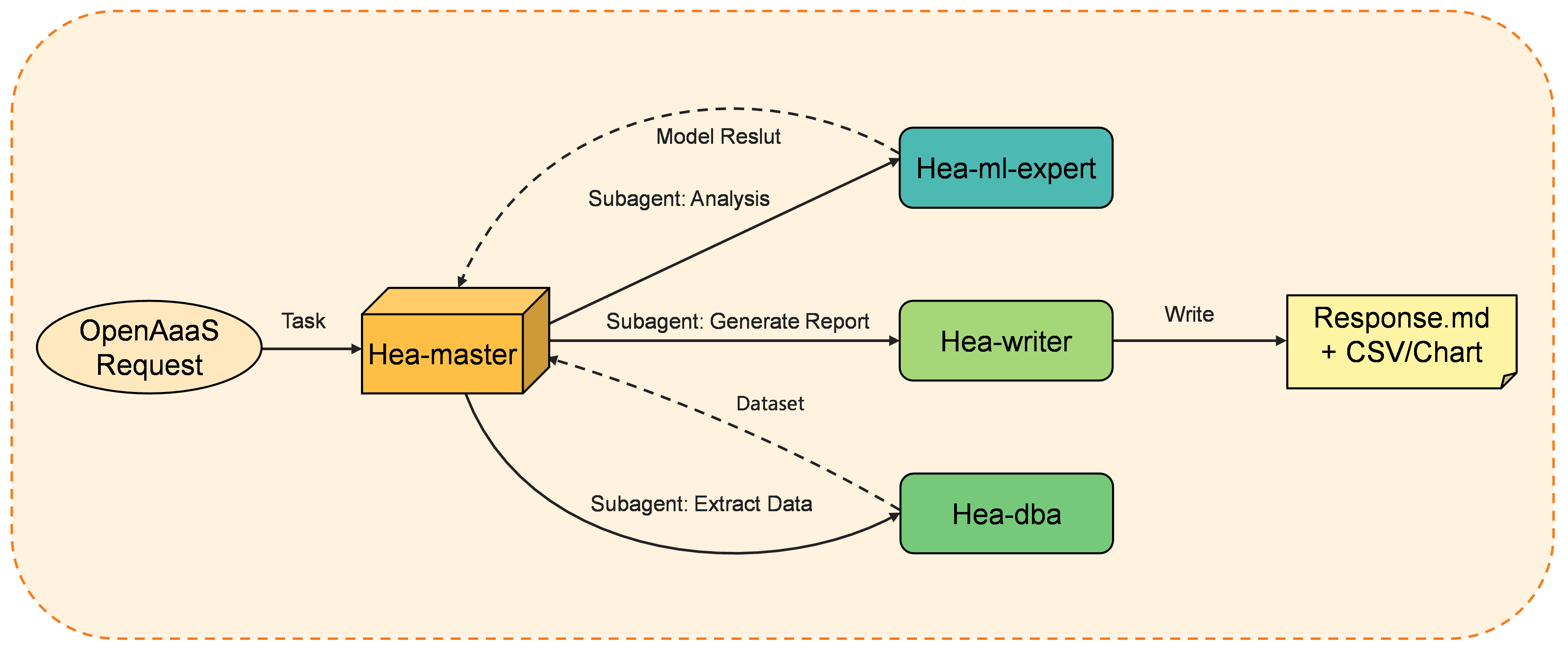}
\caption{Internal workflow of the HEA-Executor. The \texttt{hea-master} agent coordinates three domain-specific sub-agents: (i)~\texttt{hea-dba} parses natural-language requests into DuckDB-optimized SQL queries over Parquet-stored data, enforcing permutation invariance via unordered \texttt{LIST\_SORT} matching; (ii)~\texttt{hea-ml-expert} performs feature extraction, trains scikit-learn models with fixed-seed K-fold cross-validation, and outputs predictions with uncertainty quantification and feature-importance rankings; (iii)~\texttt{hea-writer} consolidates query results, ML predictions, and analytical insights into structured deliverables. Each stage preserves intermediate states for failure-mode isolation and end-to-end reproducibility.}
\label{fig:case2_workflow}
\end{figure*}

Figure~\ref{fig:case2_workflow} illustrates the internal workflow. The \texttt{hea-dba} sub-agent functions as the data-access layer. It parses natural-language task descriptions into DuckDB-optimized SQL queries and executes retrieval over the HEA database stored in columnar Parquet format. During query construction, this agent rigorously preserves materials-specific entities (elemental symbols, concentration ranges, phase labels) and employs an unordered \texttt{LIST\_SORT} matching mechanism to correctly handle the permutation invariance of six-element combinations---a constraint that naive string-matching approaches routinely violate. When the coverage of returned data is insufficient to support downstream modeling, the executor triggers bounded query reformulation (e.g., adjusting concentration tolerances, expanding elemental neighborhoods, or relaxing phase constraints) until a valid sample set meeting subsequent analytical requirements is obtained.

The \texttt{hea-ml-expert} sub-agent operates on the validated query results. It performs feature extraction and selection, trains regression or classification models using scikit-learn, and ensures reproducibility via K-fold cross-validation with a fixed random seed. Model outputs are accompanied by uncertainty quantification (prediction-interval estimates), feature-importance rankings, and standard validation metrics (coefficient of determination, mean absolute error, F1 score, or area under the ROC curve, depending on task type). This disciplined modeling stage prevents the executor from emitting uncalibrated predictions and provides the Master Agent with the quantitative diagnostics needed to assess result reliability.

Finally, the \texttt{hea-writer} sub-agent consolidates query results, machine-learning predictions, and analytical insights into structured documents---typically Markdown or HTML reports with embedded tables and figures---archiving relevant attachments to a fixed output directory to form a complete, versioned, and deliverable research product. The entire pipeline executes inside an isolated Docker container mounted with read-only database volumes, ensuring that raw data remain inaccessible even to the executor's own sub-agents except through the controlled \texttt{hea-dba} interface.

\subsection{Executor Instance and Evaluation}

To validate the practical capabilities of the HEA-Executor for complex materials-science tasks under the OpenAaaS architecture, we selected a representative design problem: \textit{Can the room-temperature plasticity of MoNbTaW refractory HEAs be optimized by adjusting compositional ratios?} This task exemplifies a typical classification--regression joint-prediction problem: first determining whether a target alloy system possesses potential room-temperature plasticity, and then quantifying the trend of plasticity indicators as a function of compositional variation. The task was submitted through the OpenAaaS client application, routed to the HEA node, and executed by the HEA-Executor without any client-side data download.

\begin{figure}[htbp]
\centering
\includegraphics[width=0.95\textwidth]{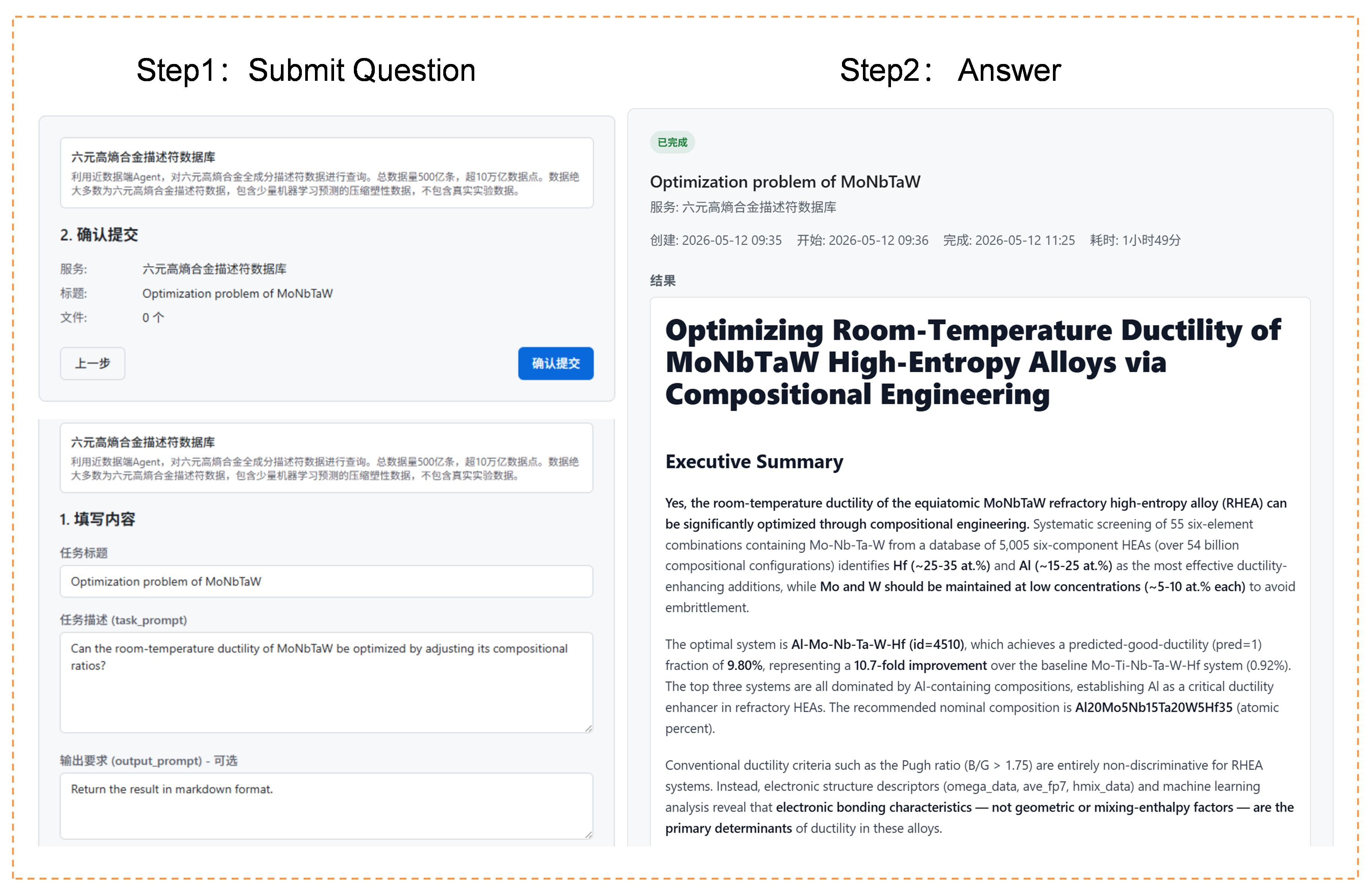}
\caption{Task submission and returned results for the HEA-Executor via the OpenAaaS client interface. The task requests identification of MoNbTaW-variant HEA systems with optimized room-temperature plasticity. The executor returns filtered candidate compositions, prediction confidence intervals, and a structured analytical report, with total returned data volume below 5~MB.}
\label{fig:case2_results}
\end{figure}

The results, summarized in Figure~\ref{fig:case2_results}, demonstrate that the HEA-Executor effectively identifies the influence trends of compositional variations on room-temperature ductility. From the full design space of 5,005 six-element systems and approximately $5.4 \times 10^{10}$ candidate configurations, the executor screened 55 MoNbTaW-containing combinations and identified the Al--Mo--Nb--Ta--W--Hf system as the optimal ductile candidate. Within this system, the proportion of highly ductile configurations reached 9.80\%, representing a 10.7-fold improvement over the conventional Ti-containing baseline system, Mo--Ti--Nb--Ta--W--Hf. Compared with conventional trial-and-error screening guided by human experience, the HEA-Executor rapidly narrows candidate regions within an enormous compositional space and provides statistically correlated optimization directions anchored to quantitative model predictions.

From an architectural perspective, this case further validates the effectiveness of the OpenAaaS near-data execution model in ultra-large-scale scientific-database scenarios. Throughout task execution, the user neither downloaded terabyte-scale databases nor directly interacted with underlying data files. The HEA-Executor completed feature analysis, model training, and inference entirely at the remote data side, returning only candidate composition ranges, prediction results, and structured reports. The actual returned data volume for this task was approximately 2.3~MB---nearly seven orders of magnitude smaller than the raw database size. This paradigm dramatically reduces demands on network bandwidth, client storage, and local computational resources while guaranteeing strict isolation of the original database structure and core data assets. The fixed orchestration overhead of the OpenAaaS task-routing layer (approximately 550~ms, as quantified in independent latency benchmarks) is negligible compared with the days of transfer time that a centralized alternative would require, and it remains constant regardless of database scale because computation is executed locally at the node.

\section{Discussion}
\label{sec:discussion}

\subsection{Positioning OpenAaaS Relative to Existing Architectures}

Table~\ref{tab:comparison_v3} positions OpenAaaS against representative systems in the related-work landscape. The comparison is organized along dimensions that matter for real-world materials-informatics deployment: whether the system supports multi-agent orchestration, whether agents can span organizational boundaries without migrating raw data, whether tools are programmatically composable, whether data sovereignty is architecturally enforced, whether the execution model supports near-data computing, and whether client compatibility is broad or framework-specific.

\begin{table}[t]
\centering
\caption{Architectural Comparison}
\label{tab:comparison_v3}
\resizebox{\columnwidth}{!}{%
\begin{tabular}{@{}lcccccc@{}}
\toprule
System & \begin{tabular}[c]{@{}c@{}}Multi-Agent\\Support\end{tabular} & \begin{tabular}[c]{@{}c@{}}Cross-Org\\Secure\end{tabular} & \begin{tabular}[c]{@{}c@{}}Tool\\Composability\end{tabular} & \begin{tabular}[c]{@{}c@{}}Data\\Sovereignty\end{tabular} & \begin{tabular}[c]{@{}c@{}}Near-Data\\Exec.\end{tabular} & \begin{tabular}[c]{@{}c@{}}Broad Client\\Compat.\end{tabular} \\
\midrule
Materials Project & $\times$ & $\times$ & API & $\times$ & $\times$ & REST \\
AutoGen & $\checkmark$ & $\times$ & Python & $\times$ & $\times$ & LangChain \\
AaaS-AN & $\checkmark$ & $\times$ & RGPS & Partial & $\times$ & Custom \\
MCP Servers & $\times$ & $\times$ & JSON-RPC & $\times$ & $\times$ & MCP Hosts \\
\textbf{OpenAaaS} & \textbf{$\checkmark$} & \textbf{$\checkmark$} & \textbf{Standard} & \textbf{$\checkmark$} & \textbf{$\checkmark$} & \textbf{MCP + Plugins} \\
\bottomrule
\end{tabular}%
}
\end{table}

Materials Project~\cite{jain2013} and similar centralized platforms offer excellent data quality and standardized APIs but are fundamentally single-agent and data-centralized. They cannot accommodate proprietary or restricted datasets, and their REST interfaces are too coarse-grained for the iterative, multi-step reasoning that LLM agents require. AutoGen~\cite{wu2023autogen} enables sophisticated multi-agent conversation patterns but assumes shared infrastructure---agents are Python objects in the same runtime, with no architectural separation between organizational domains. AaaS-AN~\cite{aaasan2025} pioneered the service-oriented agent concept but remains primarily a framework for intra-organizational deployment, with RGPS-based tool registration that requires upfront schema design.

MCP~\cite{anthropic2024mcp,hou2025} solves the client-side interoperability problem brilliantly---any MCP host can invoke any MCP server---but it is a tool-calling protocol, not an agent-network architecture. It does not define how multiple agents discover each other, how tasks are routed across nodes, how data sovereignty is maintained during multi-step workflows, or how capabilities are progressively disclosed in a network setting. OpenAaaS is designed to be complementary: it uses MCP as one of several client transport options while providing the server-side and network-level infrastructure that MCP itself does not specify.

The key differentiator of OpenAaaS is the combination of \textit{multi-agent orchestration}, \textit{cross-organizational security}, and \textit{near-data execution}. No existing system simultaneously provides all three. Materials platforms are secure and near-data but single-agent. Multi-agent frameworks are composable but not cross-organizational. AaaS protocols are interoperable but not architecturally sovereign. OpenAaaS fills this gap by treating the network itself as a composable substrate---not just for tools, but for entire agent collectives operating under distributed governance.

\subsection{Toward ``Organized Research'' with Agent Networks}

The architecture of OpenAaaS supports a broader vision that extends beyond technical interoperability: the enablement of \textit{organized research}---hierarchical, persistent, self-improving agent collectives that respect data sovereignty across institutional boundaries.

Traditional centralized platforms implicitly impose a single governance model: the platform operator controls data access, algorithm versioning, and service availability. This centralization is efficient for open science but inimical to the diverse incentive structures, regulatory requirements, and competitive pressures that characterize industrial and national-laboratory research. OpenAaaS, by contrast, enables a federation model in which each node retains full sovereignty over its data and execution policies while contributing its capabilities to a shared, discoverable network.

In this vision, a national laboratory might host high-performance computation nodes with access to supercomputing facilities; an industrial partner might host proprietary alloy databases and experimental validation services; an academic group might host open literature indices and ML model repositories. Each node independently decides which services to expose, which clients to serve, and what data to retain. The Master Agent---acting on behalf of a human researcher---composes these heterogeneous capabilities into coherent workflows without requiring any single organization to surrender control of its data.

This model also enables what we term \textit{progressive trust building}. When two institutions first connect through OpenAaaS, they might share only low-sensitivity services (e.g.,\ open literature queries). As confidence builds, they can progressively expose higher-value capabilities: proprietary descriptors, experimental synthesis services, or access to expensive characterization equipment. The service-level granularity of OpenAaaS means that trust can be calibrated service by service, not as an all-or-nothing network membership decision.

\subsection{Scalability and Limitations}

While OpenAaaS addresses a critical gap in the current landscape, several limitations and open challenges remain.

\textbf{Network scale.} The current implementation has been tested with tens of concurrent nodes. Scaling to thousands of nodes will require enhancements to the Server's task-routing algorithm, potentially incorporating distributed consensus or gossip protocols for service discovery. The SQLite backend, adequate for laboratory-scale deployments, may need to be replaced with a distributed database for very large networks.

\textbf{Semantic service discovery.} Current capability discovery relies on structured service descriptions with domain tags. As the network grows, purely tag-based discovery may become insufficient. Future work could integrate embedding-based semantic search over service descriptions, allowing Master Agents to discover relevant services through natural-language similarity rather than exact tag matching.

\textbf{Quality of service.} OpenAaaS does not currently enforce quality-of-service guarantees. A node might register a capability but provide inconsistent or incorrect results. Reputation systems, cross-validation protocols, and standardized benchmark suites are needed to establish trust in network-provided services.

\textbf{Autonomous collaboration.} The current architecture requires a human-specified Master Agent to initiate workflows. Fully autonomous inter-agent collaboration---where sub-agents proactively coordinate without a top-level orchestrator---remains an open research problem that touches on multi-agent reinforcement learning, mechanism design, and distributed consensus.

\subsection{Security Considerations}

The defense-in-depth model of OpenAaaS provides strong security for the communication and execution layers, but it does not eliminate all risks. Supply-chain attacks on container images, side-channel attacks on shared computational infrastructure, and prompt-injection attacks against LLM-based Master Agents remain concerns that must be addressed through operational practices (image signing, hardware-based isolation, prompt filtering) rather than architecture alone. The audit-trail mechanism supports forensic analysis but does not prevent exfiltration; fine-grained access control policies at the node level are ultimately the responsibility of the node operator.

\section{Conclusion}
\label{sec:conclusion}

We have presented OpenAaaS, an open Agent-as-a-Service framework for distributed materials-informatics research. By separating user-facing Master Agents from a lightweight Network Hub and near-data execution nodes, OpenAaaS enables complex multi-agent workflows that respect data sovereignty across organizational boundaries. The framework's four design propositions---data in situ, zero migration, schema-free onboarding, and near-data computing---address fundamental limitations of both centralized platforms and isolated agent systems.

Two case studies demonstrate the framework's applicability: AlphaAgent, a materials-literature analysis executor that achieves 4.66/5.0 on deep analytical questions through evidence-grounded skill composition; and an ultra-large-scale hexa-high-entropy alloy descriptor database that exposes 17.4~TB of local data as composable network services without raw-data migration. Both implementations, along with the core framework, are available as open-source software at \url{https://github.com/Wolido/OpenAaaS}.

Looking forward, three directions warrant particular attention. First, \textit{autonomous design loops}: Master Agents that not only orchestrate existing services but also propose new experimental designs, iterate based on results, and refine hypotheses without continuous human intervention. Second, \textit{cross-scale integration}: connecting atomistic simulations, continuum models, and experimental characterization through a unified agent network that spans length and time scales. Third, \textit{verifiable agent collectives}: extending OpenAaaS with formal verification and blockchain-based attestation to support high-stakes materials certification in aerospace, nuclear, and biomedical applications. These directions point toward a future in which AI-driven materials discovery is organized, persistent, and trustworthy across the global materials research enterprise.

\bibliographystyle{unsrt}
\bibliography{refs}

@article{alphaagent2026,
  title={Skill-Driven Retrieval-Augmented Generation for Intelligent Materials Science Literature Analysis},
  author={{AlphaAgent Research Team}},
  journal={Manuscript in preparation},
  year={2026}
}

@article{xiaoya2026,
  title={Trillion-Scale DataForge: Integrated Architecture for High-Throughput Materials Databases and Seamless Sharing},
  author={Xiaoya, Huang and Yu, Liu and Shuo, Shi and Yuanyuan, Zhang and Zengzeng, Liang and Miao, Zhou and Hanwei, Fu and Lei, Zheng and Peng, Kang},
  journal={Under review},
  year={2026}
}

@article{mgi2011,
  title={Materials Genome Initiative for Global Competitiveness},
  author={{National Science and Technology Council}},
  journal={Office of Science and Technology Policy},
  year={2011}
}

@article{jain2013,
  title={Commentary: The Materials Project: A Materials Genome Approach to Accelerating Materials Innovation},
  author={Jain, Anubhav and Ong, Shyue Ping and Hautier, Geoffroy and Chen, Wei and Richards, William Davidson and Dacek, Stephen and Cholia, Shreyas and Gunter, Dan and Skinner, David and Ceder, Gerbrand and Persson, Kristin A.},
  journal={APL Materials},
  volume={1},
  number={1},
  pages={011002},
  year={2013}
}

@article{curtarolo2012,
  title={AFLOW: An Automatic Framework for High-Throughput Materials Discovery},
  author={Curtarolo, Stefano and Setyawan, Wahyu and Wang, Shidong and Xue, Junkai and Yang, Kesong and Taylor, Richard H. and Nelson, Lance J. and Hart, Gus L. W. and Sanvito, Stefano and Buongiorno-Nardelli, Marco and Mingo, Natalio and Levy, Ohad},
  journal={Computational Materials Science},
  volume={58},
  pages={218--226},
  year={2012}
}

@article{kirklin2015,
  title={The Open Quantum Materials Database (OQMD): Assessing the Accuracy of DFT Formation Energies},
  author={Kirklin, Scott and Saal, James E. and Meredig, Bryce and Thompson, Alex and Doak, Jeff W. and Aykol, Muratahan and R{\"u}hl, Stephan and Wolverton, Chris},
  journal={npj Computational Materials},
  volume={1},
  pages={15010},
  year={2015}
}

@article{draxl2019,
  title={The NOMAD Laboratory: From Data Sharing to Artificial Intelligence},
  author={Draxl, Claudia and Scheffler, Matthias},
  journal={Journal of Physics: Materials},
  volume={2},
  number={3},
  pages={036001},
  year={2019}
}

@article{saal2020,
  title={Materials Data Infrastructure for the AI Era},
  author={Saal, James E. and Oses, Corey and Kirklin, Scott and Aykol, Muratahan and Wolverton, Chris},
  journal={MRS Bulletin},
  volume={45},
  number={6},
  pages={473--480},
  year={2020}
}

@article{wilkinson2016,
  title={The FAIR Guiding Principles for Scientific Data Management and Stewardship},
  author={Wilkinson, Mark D. and Dumontier, Michel and Aalbersberg, IJsbrand Jan and Appleton, Gabrielle and Axton, Myles and Baak, Arie and Blomberg, Niklas and Boiten, Jan-Willem and da Silva Santos, Luiz Bonino and Bourne, Philip E. and others},
  journal={Scientific Data},
  volume={3},
  pages={160018},
  year={2016}
}

@article{tshitoyan2019,
  title={Unsupervised Word Embeddings Capture Latent Knowledge from Materials Science Literature},
  author={Tshitoyan, Vahe and Dagdelen, John and Weston, Leigh and Dunn, Alexander and Rong, Ziqin and Kononova, Olga and Persson, Kristin A. and Ceder, Gerbrand and Jain, Anubhav},
  journal={Nature},
  volume={571},
  number={7763},
  pages={95--98},
  year={2019}
}

@article{merchant2023,
  title={Scaling Deep Learning for Materials Discovery},
  author={Merchant, Amil and Batzner, Simon and Schoenholz, Samuel S. and Aykol, Muratahan and Cheon, Gowoon and Bustamante, Joshua},
  journal={Nature},
  volume={624},
  number={7990},
  pages={80--85},
  year={2023}
}

@article{brown2020,
  title={Language Models are Few-Shot Learners},
  author={Brown, Tom B. and Mann, Benjamin and Ryder, Nick and Subbiah, Melanie and Kaplan, Jared and Dhariwal, Prafulla and Neelakantan, Arvind and Shyam, Pranav and Sastry, Girish and Askell, Amanda and others},
  journal={Advances in Neural Information Processing Systems},
  volume={33},
  pages={1877--1901},
  year={2020}
}

@article{ouyang2022,
  title={Training Language Models to Follow Instructions with Human Feedback},
  author={Ouyang, Long and Wu, Jeffrey and Jiang, Xu and Almeida, Diogo and Wainwright, Carroll and Mishkin, Pamela and Zhang, Chong and Agarwal, Sandhini and Slama, Katarina and Ray, Alex and others},
  journal={Advances in Neural Information Processing Systems},
  volume={35},
  pages={27730--27744},
  year={2022}
}

@article{achiam2023,
  title={{GPT-4} Technical Report},
  author={Achiam, Josh and Adler, Steven and Agarwal, Sandhini and Ahmad, Lama and Akkaya, Ilge and Aleman, Florencia Leoni and Almeida, Diogo and Altenschmidt, Janko and Altman, Sam and Anadkat, Shyamal and others},
  journal={arXiv preprint arXiv:2303.08774},
  year={2023}
}

@article{wu2023autogen,
  title={{AutoGen}: Enabling Next-Gen {LLM} Applications via Multi-Agent Conversation},
  author={Wu, Qingyun and Bansal, Gagan and Zhang, Jieyu and Wu, Yiran and Li, Beibin and Zhu, Erkang and Jiang, Li and Zhang, Shaokun and Liu, Jiale and Awadallah, Ahmed Hassan and others},
  journal={arXiv preprint arXiv:2308.08155},
  year={2023}
}

@article{boiko2023,
  title={Autonomous Chemical Research with Large Language Models},
  author={Boiko, Daniil A. and MacKnight, Robert and Kline, Ben and Gomes, Gabriel},
  journal={Nature},
  volume={624},
  number={7992},
  pages={570--578},
  year={2023}
}

@article{bran2024,
  title={ChemCrow: Augmenting Large-Language-Model-Based Chemical Reasoning with Specialist Tools},
  author={Bran, Andres M. and Cox, Sam and Schilter, Oliver and Baldassari, Camille and White, Andrew D. and Schwaller, Philippe},
  journal={Nature Machine Intelligence},
  volume={6},
  number={5},
  pages={525--535},
  year={2024}
}

@article{ghafarollahi2024,
  title={SciAgents: Automating Scientific Discovery through Multi-Agent Intelligent Graph Reasoning},
  author={Ghafarollahi, Alireza and Buehler, Markus J.},
  journal={arXiv preprint arXiv:2409.05556},
  year={2024}
}

@article{hong2024,
  title={MetaGPT: Meta Programming for A Multi-Agent Collaborative Framework},
  author={Hong, Sirui and Zheng, Xiang and Chen, Jonathan and Cheng, Yuhan and Wang, Ceyao and Zhang, Zili and Wang, Steven Ka Shing and Yao, Zhenqing and Wu, Bang and Zhou, Zhuorui and others},
  journal={International Conference on Learning Representations},
  year={2024}
}

@article{li2023camel,
  title={{CAMEL}: Communicative Agents for ``Mind'' Exploration of Large Language Model Society},
  author={Li, Guohao and Hammoud, Hasan Abed Al Kader and Itani, Hadi and Khizbullin, Dmitrii and Ghanem, Bernard},
  journal={Advances in Neural Information Processing Systems},
  volume={36},
  year={2023}
}

@article{lewis2020retrieval,
  title={Retrieval-Augmented Generation for Knowledge-Intensive {NLP} Tasks},
  author={Lewis, Patrick and Perez, Ethan and Piktus, Aleksandra and Petroni, Fabio and Karpukhin, Vladimir and Goyal, Naman and Kuttler, Heinrich and Lewis, Mike and Yih, Wen-tau and Rockt{\"a}schel, Tim and Riedel, Sebastian and Kiela, Douwe},
  journal={Advances in Neural Information Processing Systems},
  volume={33},
  pages={9459--9474},
  year={2020}
}

@article{izacard2021leveraging,
  title={Leveraging Passage Retrieval with Generative Models for Open Domain Question Answering},
  author={Izacard, Gautier and Grave, Edouard},
  journal={Proceedings of the 16th Conference of the European Chapter of the Association for Computational Linguistics},
  pages={874--880},
  year={2021}
}

@article{anthropic2024mcp,
  title={Model Context Protocol},
  author={{Anthropic}},
  journal={Anthropic Technical Documentation},
  year={2024},
  url={https://modelcontextprotocol.io}
}

@article{liu2024matclaw,
  title={MatClaw: An Autonomous Code-First {LLM} Agent for End-to-End Materials Exploration},
  author={Liu, Zihan and Zhang, Yong and Wang, Chenxi and others},
  journal={arXiv preprint arXiv:2604.02688},
  year={2026}
}

@article{blockchain2024,
  title={Blockchain Technology for Big-data Sharing in Material Genome Engineering},
  author={Chen, Xingyu and others},
  journal={Journal of Materials Informatics},
  year={2024}
}

@article{aaasan2025,
  title={Agent-as-a-Service based on Agent Network},
  author={Li, Wei and Zhang, Jie and others},
  journal={arXiv preprint arXiv:2505.08446},
  year={2025}
}

@article{ren2025,
  title={A Survey on LLM-based Scientific Agents},
  author={Ren, Xiang and others},
  journal={arXiv preprint arXiv:2503.24047},
  year={2025}
}

@article{agarwal2024,
  title={LitLLM: A Toolkit for Literature Review with Large Language Models},
  author={Agarwal, Shivam and others},
  journal={arXiv preprint arXiv:2402.01788},
  year={2024}
}

@article{lin2024,
  title={Paper Copilot: A Personalized Research Assistant},
  author={Lin, Yijia and others},
  journal={arXiv preprint arXiv:2403.12345},
  year={2024}
}

@article{lu2024,
  title={The {AI} Scientist: Towards Fully Automated Open-ended Scientific Discovery},
  author={Lu, Chris and others},
  journal={arXiv preprint arXiv:2408.06292},
  year={2024}
}

@article{yamada2025,
  title={The {AI} Scientist-v2: Workshop-Ready Automated Research},
  author={Yamada, Yutaro and others},
  journal={ICLR Workshop on Machine Learning for Materials},
  year={2025}
}

@article{gottweis2025,
  title={{AI} Co-Scientist: A Multi-Agent System for Scientific Discovery},
  author={Gottweis, Jonas and others},
  journal={Google DeepMind Technical Report},
  year={2025}
}

@article{vriza2026,
  title={Multi-Agent Frameworks for Atomistic Simulations},
  author={Vriza, Aikaterini and others},
  journal={npj Computational Materials},
  year={2026}
}

@article{sanchez2018,
  title={Inverse Molecular Design using Machine Learning: Generative Models for Matter Engineering},
  author={S{\'a}nchez-Lengeling, Benjamin and Aspuru-Guzik, Al{\'a}n},
  journal={Science},
  volume={361},
  number={6400},
  pages={360--365},
  year={2018}
}

@article{ramakrishna2019,
  title={Materials Informatics: Status, Challenges and Perspectives},
  author={Ramakrishna, Seeram and Zhang, Tao and Lu, Wen Feng and others},
  journal={Journal of Intelligent Manufacturing},
  volume={30},
  pages={2307--2326},
  year={2019}
}

@article{zhang2024honeycomb,
  title={HoneyComb: Flexible LLM-Based Agents for Materials Science with Domain Knowledge Bases},
  author={Zhang, Hao and others},
  journal={Nature Communications},
  year={2024}
}

@article{antunes2024,
  title={Crysta{LLM}: An Autoregressive LLM for the Versatile Generation of Crystal Structures},
  author={Antunes, L. M. and others},
  journal={Nature Communications},
  year={2024}
}

@article{hasan2025,
  title={Security Threats in Model Context Protocol: A Comprehensive Analysis},
  author={Hasan, Mohammad and others},
  journal={arXiv preprint arXiv:2503.23278},
  year={2025}
}

@article{hou2025,
  title={MCP Server Landscape and Maintainability Analysis},
  author={Hou, Yuxin and others},
  journal={arXiv preprint arXiv:2506.13538},
  year={2025}
}

@article{wei2022,
  title={Emergent Abilities of Large Language Models},
  author={Wei, Jason and Tay, Yi and Bommasani, Rishi and Raffel, Colin and Zoph, Barret and Borgeaud, S{\'e}bastien and Yogatama, Dani and Bosma, Maarten and Zhou, Denny and Metzler, Donald and Chi, Ed H. and Hashimoto, Tatsunori and Vinyals, Oriol and Liang, Percy and Dean, Jeff and Fedus, William},
  journal={Transactions on Machine Learning Research},
  year={2022}
}

@article{tang2024,
  title={Towards Scientific Intelligence: A Survey of LLM-based Scientific Agents},
  author={Tang, Jing and others},
  journal={arXiv preprint arXiv:2503.24047},
  year={2025}
}

@article{sears2024,
  title={{SEARS}: A Lightweight {FAIR} Platform for Multi-Lab Materials Collaboration},
  author={Sears, Matthew and others},
  journal={Materials Discovery},
  volume={3},
  pages={100013},
  year={2024}
}

@article{ghafarollahi2025pnas,
  title={Physics-Aware Multimodal Multi-Agent Systems for Alloy Design and Discovery},
  author={Ghafarollahi, Alireza and Buehler, Markus J.},
  journal={Proceedings of the National Academy of Sciences},
  year={2025},
  publisher={National Academy of Sciences}
}

@article{miret2025nmi,
  title={Large Language Models for Scientific Discovery: Opportunities and Challenges},
  author={Miret, Santiago and Krishnan, Arvind},
  journal={Nature Machine Intelligence},
  year={2025},
  publisher={Nature Publishing Group}
}

@article{wei2025agentic,
  title={Agentic {AI} for Scientific Discovery: A Survey of Autonomous Research Systems},
  author={Wei, Jason and others},
  journal={arXiv preprint arXiv:2501.03200},
  year={2025}
}

@article{gridach2025survey,
  title={A Comprehensive Survey of Multi-Agent Systems for Scientific Discovery},
  author={Gridach, Mourad and others},
  journal={arXiv preprint arXiv:2502.01000},
  year={2025}
}

@article{pham2025chemgraph,
  title={{ChemGraph}: A Graph-Based Multi-Agent Framework for Autonomous Chemical Discovery},
  author={Pham, Trang and others},
  journal={Digital Discovery},
  year={2025},
  publisher={Royal Society of Chemistry}
}

@article{wang2025dreams,
  title={{DREAMS}: A Density Functional Theory Based Research Engine for Agentic Materials Simulation},
  author={Wang, Yining and others},
  journal={npj Computational Materials},
  year={2025},
  publisher={Nature Publishing Group}
}

@article{feng2026internagent,
  title={{InternAgent-1.5}: A Unified Agentic Framework for Long-Horizon Autonomous Scientific Discovery},
  author={Feng, Zekun and others},
  journal={arXiv preprint arXiv:2506.00000},
  year={2026}
}

@article{horton2025natmat,
  title={The Materials Project: Accelerating Materials Design through Open-Access Data and Tools},
  author={Horton, Matthew and others},
  journal={Nature Materials},
  year={2025},
  publisher={Nature Publishing Group}
}

@article{jiang2025npj,
  title={Large Language Models in Materials Science: From Property Prediction to Autonomous Discovery},
  author={Jiang, Shengdong and others},
  journal={npj Computational Materials},
  year={2025},
  publisher={Nature Publishing Group}
}

\end{document}